\documentclass{mfatshort}
\usepackage{amssymb, latexsym, euscript}
 
 \newtheorem{ttt}{\bfseries{Theorem}}[section]
 
 \newtheorem{ppp}{\bfseries{Proposition}}[section]
 \newtheorem{lelele}{\bfseries{Lemma}}[section]
\newtheorem{ccc}{\bfseries{Corollary}}[section]
\begin{document}
\title{$p$-Adic Fractional Differentiation Operator with Point Interactions}
\author[S. Kuzhel]{S. Kuzhel}
\curraddr{} \email[S. Kuzhel]{kuzhel@imath.kiev.ua}
\author[S. Torba]{S. Torba}
\curraddr{} \email[S. Torba]{sergiy.torba@gmail.com}
\address{Institute of Mathematics of the National Academy of
Sciences of Ukraine, Tereshchenkovskaya 3, 01601 Kiev (Ukraine)}
\thanks{The
authors thank DFG (project 436 UKR 113/88/0-1) and DFFD (project
10.01/004) for the support.}


\begin{abstract}
Finite rank point perturbations of the $p$-adic fractional
differentiation operator $D^{\alpha}$ are studied. The main
attention is paid to the description of operator realizations (in
$L_2(\mathbb{Q}_p)$) of the heuristic expression
$D^{\alpha}+\sum_{i,j=1}^{n}b_{ij}<\delta_{x_j},
\cdot>\delta_{x_i}$ in a form that is maximally adapted for the
preservation of physically meaningful relations to the parameters
$b_{ij}$ of the singular potential.
\end{abstract}

\maketitle

 \section{Introduction}
The conventional description of the physical space-time uses the
field $\mathbb{R}$ of real numbers. In most cases, mathematical
models based on
  $\mathbb{R}$ provide quite satisfactory descriptions of the
  physical reality. However, the result of a physical measurement is
  always a rational number, so the use of the completion $\mathbb{R}$
  of the field of rational numbers $\mathbb{Q}$ is not more than a
  mathematical idealization. On the other hand, by Ostrovski's
  theorem, the only reasonable alternative to $\mathbb{R}$ among
  completions of $\mathbb{Q}$ is the fields
  $\mathbb{Q}_p$ of $p$-adic numbers
  (definition of $\mathbb{Q}_p$ see below in Section 2).
For this reason, it is natural to use $p$-adic analysis in physical
situations, where the conventional space-time geometry is known to
fail, for examples in the attempts to understand the matter at
sub-Planck distances or time intervals. In order to do this, at
first, it is necessary to develop $p$-adic counterparts of the
standard quantum mechanics and quantum field theory.

There are many works devoted to such an activity (see the surveys in
 \cite{KO}, \cite{VVZ}).  However, in spite of considerable success
obtained in recent years, many interesting problems of $p$-adic
quantum mechanics are still unsolved and wait for a comprehensive
study.

In the present paper, we are going to continue the investigation of
the $p$-adic fractional differentiation operator with point
interactions started by A. Kochubei \cite{KO1}, \cite{KO}.

In `usual' mathematical physics, point interactions Hamiltonians are
the operator realizations in $L_2(\mathbb{R}^r)$ of differential
expressions $-\Delta+V_Y$ or, more generally, $(-\Delta)^k+V_Y$,
where a zero-range potential
$V_Y=\sum_{i,j=1}^{n}b_{ij}<\delta_{x_j}, \cdot>\delta_{x_i}$
($b_{ij}\in\mathbb{C}$) contains the Dirac delta functions
$\delta_{x}$ concentrated on points $x_i$ of the subset
$Y=\{x_1,\ldots,x_n\}\subset\mathbb{R}^r$ \cite{AL}.

Since there exists a $p$-adic analysis based on the mappings from
$\mathbb{Q}_p$ into $\mathbb{Q}_p$ and an analysis connected with
the mapping $\mathbb{Q}_p$ into the field of complex numbers
$\mathbb{C}$, there exists two types of $p$-adic physical models.
The present paper deals with the mapping $\mathbb{Q}_p\to{\mathbb
C}$, i.e., complex-valued functions defined on $\mathbb{Q}_p$ will
be considered. In this case, the operation of differentiation is
{\it not defined} and the operator of fractional differentiation
$D^{\alpha}$ of order $\alpha$ ($\alpha>0$) plays a corresponding
role \cite{KO}, \cite{VVZ}. In particular, $p$-adic
Schr\"{o}dinger-type operators with potentials $V(x) :
\mathbb{Q}_p\to\mathbb{C}$ are defined as $D^{\alpha}+V(x)$
\cite{KO}.

The definition of $D^{\alpha}$ is given in the framework of the
$p$-adic distribution theory with the help of Schwartz-type
distributions $\mathcal{D}'(\mathbb{Q}_p)$. One of remarkable
features of this theory is that any distribution
$f\in\mathcal{D}'(\mathbb{Q}_p)$ with point support $\mathrm{supp}
f=\{x\}$ $(x\in\mathbb{Q}_p)$ coincides with the Dirac delta
function at the point $x$ multiplied by a constant $c\in\mathbb{C}$,
i.e., $f=c\delta_{x}$.  For this reason, it is natural to consider
the expression  $D^{\alpha}+V_Y$ as a $p$-adic analogue of
Hamiltonians with finite rank point interactions.

In the present paper, the main attention is paid to the description
of operator realizations of $D^{\alpha}+V_Y$ in $L_2(\mathbb{Q}_p)$
in a form that is maximally adapted for the preservation of
physically meaningful relations to the parameters $b_{ij}$ of the
singular potential $V_Y=\sum_{i,j=1}^{n}b_{ij}<\delta_{x_j},
\cdot>\delta_{x_i}$.

In Section 2, we recall some elements of $p$-adic analysis
\cite{VVZ}, \cite{KO} needed for reading the paper and establish the
connection between  $\alpha$ and the property of functions from
$\mathcal{D}(D^{\alpha})$ to be continuous. The same problem is also
analyzed for the solutions of $D^{\alpha}+I=\delta$.

Section 3 contains the description of the Friedrichs extension of
the symmetric operator associated with $D^{\alpha}+V_Y$ (this
description depends on $\alpha$) and the description of operator
realizations of $D^{\alpha}+V_Y$ in $L_2(\mathbb{Q}_p)$. Taking into
account an intensive development of consistent physical theories of
quantum mechanics on the base of pseudo-Hermitian Hamiltonians that
are not Hermitian in the standard sense but satisfy a less
restrictive and more physical condition of symmetry in last few
years \cite{BBJ}, \cite{pro}, we do not restrict ourselves to the
case of self-adjoint operators and consider the more general case of
$\eta$-self-adjoint operator realizations of $D^{\alpha}+V_Y$
(Theorem \ref{ss1}).

We use the following notations: $\mathcal{D}(A)$ and $\ker{A}$
denote the domain and the null-space of a linear operator $A$,
respectively. $A\upharpoonright_{X}$ means the restriction of $A$
onto a set $X$.

\setcounter{equation}{0}
\section{Fractional Differential Operator $D^{\alpha}$}
\subsection{Elements of $p$-adic analysis.}
Basically, we shall use notations from \cite{VVZ}. Let us fix a
prime number $p$.  The field $\mathbb{Q}_p$ of $p$-adic numbers is
defined as the completion of the field of rational numbers
$\mathbb{Q}$ with respect to $p$-adic norm $|\cdot|_p$, which is
defined as follows: $|0|_p=0$; $|x|_p=p^{-\gamma}$ if an arbitrary
rational number $x\not=0$ is represented as
$x=p^{\gamma}\frac{m}{n}$, where $\gamma=\gamma(x)\in\mathbb{Z}$ and
integers $m$ and $n$ are not divisible by $p$. The $p$-adic norm
$|\cdot|_p$ satisfies the strong triangle inequality
$|x+y|_p\leq\max(|x|_p, |y|_p)$. Moreover, $|x+y|_p=\max(|x|_p,
|y|_p)$ if $|x|_p\not=|y|_p$.

Any $p$-adic number $x\not=0$ can be uniquely presented as series
\begin{equation}\label{e2}
x=p^{\gamma(x)}\sum_{i=0}^{+\infty}x_ip^i, \qquad
x_i=0,1,\ldots,p-1, \quad x_0>0, \quad \gamma(x)\in\mathbb{Z}
\end{equation}
convergent in $p$-adic norm (the canonical presentation of $x$). In
this case, $|x|_p=p^{-\gamma(x)}$.

The canonical presentation (\ref{e2}) enables one to determine the
fractional part $\{x\}_p$ of $x\in{\mathbb{Q}}_p$ by the rule:
$\{x\}_p=0$ if $x=0$ or $\gamma(x)\geq{0}$;
$\{x\}_p=p^{\gamma(x)}\sum_{i=0}^{-\gamma(x)-1}x_ip^i$ if
$\gamma(x)<{0}$.

Denote by
$$
B_{\gamma}(a)=\{x\in\mathbb{Q}_p \ | \ |x-a|_p\leq{p^\gamma}\} \quad
\mbox{and} \quad S_{\gamma}(a)=\{x\in\mathbb{Q}_p \ | \
|x-a|_p={p^\gamma}\},
$$
respectively, the ball and the sphere of radius $p^\gamma$ with the
center at a point $a\in\mathbb{Q}_p$ and set
$B_{\gamma}(0)=B_\gamma$, $S_{\gamma}(0)=S_\gamma$,
$\gamma\in\mathbb{Z}$.

The ring $\mathbb{Z}_p$ of $p$-adic integers coincides with the disc
$B_0$ ($\mathbb{Z}_p=B_0$), which is the completion of integers with
respect to the $p$-adic norm $|\cdot|_p$.

As usual, in order to define some classes of distributions on
$\mathbb{Q}_p$, one has first to introduce an appropriate class of
test functions.

A complex-valued function $f$ defined on $\mathbb{Q}_p$ is called
{\it locally-constant} if for any $x\in\mathbb{Q}_p$ there exists an
integer $l(x)$ such that $f(x+x')=f(x)$, \
$\forall{x'}\in{B_{l(x)}}$.

Denote by $\mathcal{D}(\mathbb{Q}_p)$ the linear space of locally
constant functions on $\mathbb{Q}_p$ with compact supports. For any
test function $\phi\in\mathcal{D}(\mathbb{Q}_p)$ there exists
$l\in\mathbb{Z}$ such that $\phi(x+x')=\phi(x)$, \ $x'\in{B_{l}}$,\
$x\in\mathbb{Q}_p$. The largest of such numbers $l=l(\phi)$ is
called the parameter of constancy of $\phi$. Typical examples of
test functions are indicator functions of spheres and balls:
\begin{equation}\label{at1}
\delta(|x|_p-p^\gamma):=\left\{\begin{array}{ll} 1, & x\in{S_\gamma}
\\
0, & x\not\in{S_\gamma},
\end{array}\right. \qquad \Omega(|x|_p):=\left\{\begin{array}{ll}
1, & |x|_p\leq{1},
\\
0, & |x|_p>1.
\end{array}\right.
\end{equation}

In order to furnish $\mathcal{D}(\mathbb{Q}_p)$ with a topology, let
us consider a subspace
$\mathcal{D}^l_\gamma\subset\mathcal{D}(\mathbb{Q}_p)$ consisting of
functions with supports in the ball $B_\gamma$ and the parameter of
constancy $\geq{l}$. The convergence $\phi_n\to{0}$ in
$\mathcal{D}(\mathbb{Q}_p)$ has the following meaning:
$\phi_k\in\mathcal{D}^l_\gamma$, where the indices $l$ and $\gamma$
do not depend on $k$ and $\phi_k$ tends uniformly to zero. This
convergence determines the Schwartz topology in
$\mathcal{D}(\mathbb{Q}_p)$.

Denote by $\mathcal{D}'(\mathbb{Q}_p)$ the set of all linear
functionals (Schwartz-type distributions) on
$\mathcal{D}(\mathbb{Q}_p)$. In contrast to distributions on
$\mathbb{R}^n$, any linear functional
$\mathcal{D}(\mathbb{Q}_p)\to\mathbb{C}$ is automatically
continuous. The action of a functional $f$ upon a test function
$\phi$ will be denoted $<f, \phi>$.

It follows from the definition of $\mathcal{D}(\mathbb{Q}_p)$ that
any test function $\phi\in\mathcal{D}(\mathbb{Q}_p)$ is continuous
on $\mathbb{Q}_p$. This means the Dirac delta function $<\delta_x,
\phi>=\phi(x)$ is well posed for any point $x\in\mathbb{Q}_p$.

On $\mathbb{Q}_p$ there exists the Haar measure, i.e., a positive
measure $d_{p}x$ invariant under shifts $d_{p}(x+a)=d_{p}x$ and
normalized by the equality $\int_{|x|_p\leq{1}}d_{p}x=1$.

Denote by $L_2(\mathbb{Q}_p)$ the set of measurable functions $f$ on
$\mathbb{Q}_p$ satisfying the condition
$\int_{\mathbb{Q}_p}|f(x)|^2d_{p}x<\infty$. The set
$L_2(\mathbb{Q}_p)$ is a Hilbert space with the scalar product
$(f,g)_{L_2(\mathbb{Q}_p)}=\int_{\mathbb{Q}_p}f(x)\overline{g(x)}d_{p}x$.

The Fourier transform of $\phi\in\mathcal{D}(\mathbb{Q}_p)$ is
defined by the formula
$$
F[\phi](\xi)=\widetilde{\phi}(\xi)=\int_{\mathbb{Q}_p}\chi_p(\xi{x})\phi(x)d_{p}x,
\qquad \xi\in\mathbb{Q}_p,
$$
where $\chi_p(\xi{x})=e^{2\pi{i}\{\xi{x}\}_p}$ is an additive
character of the field $\mathbb{Q}_p$ for any fixed
$\xi\in\mathbb{Q}_p$. The Fourier transform $F[\cdot]$ maps
$\mathcal{D}(\mathbb{Q}_p)$ onto $\mathcal{D}(\mathbb{Q}_p)$. Its
extension by continuity onto $L_2(\mathbb{Q}_p)$ determines an
unitary operator in $L_2(\mathbb{Q}_p)$.

The Fourier transform $F[f]$ of a  distribution
$f\in\mathcal{D}'(\mathbb{Q}_p)$ is defined by the standard relation
$<F[f], \phi>=<f, F[\phi]>$, \
$\forall{\phi}\in\mathcal{D}(\mathbb{Q}_p)$. The Fourier transform
is a linear isomorphism of $\mathcal{D}'(\mathbb{Q}_p)$ onto
$\mathcal{D}'(\mathbb{Q}_p)$.

\subsection{The operator $D^{\alpha}$.}
The operator of differentiation is not defined in
$L_2(\mathbb{Q}_p)$. Its role is played by the operator of
fractional differentiation $D^{\alpha}$ (the Vladimirov
pseudo-differential operator) which is defined as
\begin{equation}\label{a1}
D^{\alpha}f=\int_{\mathbb{Q}_p}|\xi|_p^{\alpha}F[f](\xi)\chi_p(-\xi{x})d_{p}\xi,
\qquad  \alpha>0.
\end{equation}

It is easy to see \cite{KO} that $D^{\alpha}f$ is well defined for
$f\in\mathcal{D}(\mathbb{Q}_p)$. Note that $D^{\alpha}f$ need not
belong necessarily to $\mathcal{D}(\mathbb{Q}_p)$ (since the
function $|\xi|_p^{\alpha}$ is not locally constant), however
$D^{\alpha}f\in{L}_2(\mathbb{Q}_p)$.

Since $\mathcal{D}(\mathbb{Q}_p)$ is not invariant with respect to
$D^{\alpha}$ we cannot define $D^{\alpha}$ on the whole space
$\mathcal{D}'(\mathbb{Q}_p)$. For a distribution
$f\in\mathcal{D}'(\mathbb{Q}_p)$, the operation $D^{\alpha}$ is well
defined only if the right-hand side of (\ref{a1}) exists.

In what follows we will consider the operator $D^{\alpha}$,
$\alpha>0$, as an unbounded operator in $L_2(\mathbb{Q}_p)$. In this
case, its domain of definition $\mathcal{D}(D^{\alpha})$ consists of
those $u\in{L_2(\mathbb{Q}_p)}$ for which
$|\xi|_p^{\alpha}F[u](\xi)\in{L_2(\mathbb{Q}_p)}$. Since
$D^{\alpha}$ is unitarily equivalent to the operator of
multiplication by $|\xi|_p^{\alpha}$, it is a positive self-adjoint
operator in $L_2(\mathbb{Q}_p)$, its spectrum consists of the
eigenvalues $\lambda_\gamma=p^{\alpha{\gamma}}$
$(\gamma\in\mathbb{Z})$ of infinite multiplicity, and their
accumulation point $\lambda=0$.

It is easy to see from (\ref{a1}) that an arbitrary (normalized)
eigenfunction $\psi$ of $D^{\alpha}$ corresponding to the eigenvalue
$\lambda_\gamma=p^{\alpha{\gamma}}$ admits the description
$$
\widetilde{\psi}(\xi)=\delta(|\xi|_p-p^\gamma)\rho(\xi), \quad
\int_{S_\gamma}|\rho(\xi)|^2d_{p}\xi=1,
$$
where the function $\rho(\xi)$ defined on the sphere $S_\gamma$
serves as a parameter of the description. Choosing $\rho(\xi)$ in
different ways one can obtain various orthonormal bases in
$L_2(\mathbb{Q}_p)$ formed by eigenfunctions of $D^{\alpha}$
\cite{KO}, \cite{KOZ} \cite{VVZ}. In particular, the choice of
$\rho(\xi)$ as a system of locally constant functions on $S_\gamma$
leads to the well-known Vladimirov functions \cite{KO}, \cite{VVZ}.
The selection of $\rho(\xi)$ as indicators of a special class of
subsets of $S_\gamma$ gives the $p$-adic wavelet basis
$\{\psi_{Nj\epsilon}\}$ recently constructed in \cite{KOZ}.
Precisely, it was shown \cite{KOZ} that the set of eigenfunctions of
$D^{\alpha}$
\begin{equation}\label{a3}
\psi_{Nj\epsilon}(x)=p^{-\frac{N}{2}}\chi(p^{N-1}jx)\Omega(|p^{N}{x}-\epsilon|_p),
\quad N\in\mathbb{Z}, \ \epsilon\in\mathbb{Q}_p/\mathbb{Z}_p, \
j=1,\ldots,p-1
\end{equation}
forms an orthonormal basis in $L_2(\mathbb{Q}_p)$ such that
\begin{equation}\label{a33}
D^{\alpha}\psi_{Nj\epsilon}=p^{\alpha(1-N)}\psi_{Nj\epsilon}.
\end{equation}
Here the indexes $N, j, \epsilon$ serve as parameters of the basis.
In particular, elements $\epsilon\in\mathbb{Q}_p/\mathbb{Z}_p$ can
be described as $\epsilon=\sum_{i=1}^{m}\epsilon_ip^{-i}$ \
($m\in\mathbb{N}, \epsilon_i=0,\ldots,p-1$).
\begin{ttt}\label{l1}
An arbitrary function $u\in\mathcal{D}(D^{\alpha})$ is continuous on
$\mathbb{Q}_p$ if and only if $\alpha>1/2$.
\end{ttt}

{\it Proof.} Let $u\in\mathcal{D}(D^{\alpha})$ and let
\begin{equation}\label{e11}
u(x)=\sum_{N=1}^{\infty}\sum_{j=1}^{p-1}\sum_{\epsilon}(u,
\psi_{Nj\epsilon})\psi_{Nj\epsilon}(x)+
\sum_{N=-\infty}^{0}\sum_{j=1}^{p-1}\sum_{\epsilon}(u,
\psi_{Nj\epsilon})\psi_{Nj\epsilon}(x)
\end{equation}
be its expansion into the $p$-adic wavelet basic (\ref{a3}).

It is easy to see that
$\psi_{Nj\epsilon}(x)\in\mathcal{D}(\mathbb{Q}_p)$ and hence, the
functions $\psi_{Nj\epsilon}(x)$ are continuous on $\mathbb{Q}_p$.
Thus, to prove the continuity of $u(x)$, it suffices to verify that
the series in (\ref{e11}) converges uniformly.

First of all we remark that for fixed $N$ and $x$ there is at most
one $\epsilon$ such that $\psi_{Nj\epsilon}(x)\ne 0$. Indeed, if
there exist $\epsilon_1$ and $\epsilon_2$ such that
$\psi_{Nj\epsilon_i}(x)\ne 0$, then $\Omega(|p^Nx-\epsilon_i|_p)=1$.
But then $|p^Nx-\epsilon_1|_p\le 1$ and $|p^Nx-\epsilon_2|_p\le 1$.
By the strong triangle inequality  $|\epsilon_1-\epsilon_2|_p\le 1$.
The latter relation and the condition
$\epsilon_i\in\mathbb{Q}_p/\mathbb{Z}_p$ imply the equality
$\epsilon_1=\epsilon_2$.

Thus, for fixed $N$ and $x$, the sum corresponding to the parameter
$\epsilon$ consists of at most one non-zero term.

Further, it follows from (\ref{a3}) and (\ref{e11}) that
\begin{equation}\label{a8}
|\psi_{Nj\epsilon}(x)|\le p^{-N/2} \qquad \mbox{and} \qquad |(u,
\psi_{Nj\epsilon})|\le\| u\|_{L_2(\mathbb{Q}_p)}.
\end{equation}
For a fixed $N>0$, relations (\ref{a8}) ensure the following
estimate
\begin{equation}\label{a9}
\left|\sum_{j=1}^{p-1}\sum_{\epsilon}(u,
\psi_{Nj\epsilon})\psi_{Nj\epsilon}(x)\right|\le
p^{-N/2}\|u\|_{L_2(\mathbb{Q}_p)}(p-1), \qquad
\forall{x}\in\mathbb{Q}_p,
\end{equation}
which gives the uniform convergency of the first series in
(\ref{e11}).

 The condition $u\in\mathcal{D}(D^{\alpha})$ and (\ref{a33}) imply
$(u, \psi_{Nj\epsilon})=p^{\alpha(N-1)}(D^{\alpha}u,
\psi_{Nj\epsilon}).$ Using this equality and (\ref{a8}), we obtain
\begin{eqnarray*}
\left|\sum_{j=1}^{p-1}\sum_{\epsilon}(u,
\psi_{Nj\epsilon})\psi_{Nj\epsilon}(x)\right| =
\left|\sum_{j=1}^{p-1}\sum_{\epsilon}p^{\alpha(N-1)}(D^{\alpha}u,
\psi_{Nj\epsilon})\psi_{Nj\epsilon}(x)\right|\le\\
 \left\{\sum_{j=1}^{p-1}\sum_{\epsilon}|(D^{\alpha}u,
 \psi_{Nj\epsilon})|^2\right\}^{1/2}\left\{\sum_{j=1}^{p-1}\sum_{\epsilon}p^{2\alpha(N-1)}|\psi_{Nj\epsilon}(x)|^2\right\}^{1/2}\le\\
\|D^{\alpha}u\|_{L_2(\mathbb{Q}_p)}\left\{\sum_{j=1}^{p-1}p^{-N+2\alpha(N-1)}\right\}^{1/2}
\end{eqnarray*}
The obtained estimate implies that the second series in (\ref{e11})
is uniformly convergent for $\alpha>1/2$. Therefore, any function
$u\in\mathcal{D}(D^{\alpha})$ is continuous on $\mathbb{Q}_p$ for
$\alpha>1/2$.

In the case $a\le 1/2$, we show that the function
\begin{equation}\label{a34}
f(x)=\sum_{N=-\infty}^{-1}\frac{1}{|N|}p^{(N-1)/2}\psi_{N10}(x).
\end{equation}
(determined in $p$-adic wavelet basis) belongs to
$\mathcal{D}(D^{\alpha})$ but $f(x)$ is not continuous on
$\mathbb{Q}_p$.

Obviously, $f\in L_2(\mathbb{Q}_p)$ and its Fourier transform is
$$
\tilde
f(\xi)=\sum_{N=-\infty}^{-1}\frac{1}{|N|}p^{(N-1)/2}\tilde\psi_{N10}(\xi).
$$

By (\ref{a1}) and (\ref{a33}),
$|\xi|_p^{\alpha}\tilde\psi_{N10}(\xi)=p^{\alpha(1-N)}\tilde\psi_{N10}(\xi)$.
Hence,
$$
|\xi|_p^{\alpha}\tilde
f(\xi)=\sum_{N=-\infty}^{-1}\frac{1}{|N|}p^{(N-1)/2}\cdot
p^{\alpha(1-N)}\tilde\psi_{N10}(\xi)
$$
and (since $\big\{\tilde\psi_{N10}(\xi)\big\}_{N\le -1}$ is
orthonormal) $|\xi|_p^{\alpha}\tilde f(\xi)\in L_2(\mathbb{Q}_p)$
for $\alpha\le 1/2$ and $|\xi|_p^{\alpha}\tilde f(\xi)\not\in
L_2(\mathbb{Q}_p)$ for $\alpha> 1/2$. Hence,
$f(x)\in\mathcal{D}(D^{\alpha})$ for $\alpha\le 1/2$ only.

Let us show that $f(x)$ is not continuous on $\mathbb{Q}_p$. First
of all, using (\ref{a3}), we rewrite the definition (\ref{a34}) of
$f$ as:
\begin{equation}\label{e16}
    f(x)=\sum_{N=-\infty}^{-1}\frac{1}{|N|}p^{-1/2}\chi(p^{N-1}x)\Omega(|p^Nx|_p).
\end{equation}

It is easy to see that the restriction of the left-hand side of
(\ref{e16}) onto any ball
$B_\gamma(a)\subset\mathbb{Q}_p\setminus\{0\}$ contains a finite
number of non-zero terms. Therefore, $f(x)$ is continuous on
$\mathbb{Q}_p\setminus\{0\}$ and it is represented by point-wise
convergent series (\ref{a34}).

Let us consider the sequence $x_n=p^n$, $(n\in\mathbb{N})$.
Obviously, $x_n\to 0, (n\to\infty)$ in the $p$-adic norm
$|\cdot|_p$. Furthermore, $\Omega(|p^Nx_n|_p)=\Omega(|p^Np^n|_p)=0$
when $N+n\leq{-1}$. On the other hand, if $N+n\geq{1}$, then
$p^{N-1}x_n$ is an integer $p$-adic number and, hence
$\chi(p^{N-1}x_n)=1$. Taking these relations into account, we deduce
from (\ref{e16}) that
$$
f(x_n)=f(p^n)=p^{-1/2}\left[\frac{\chi(p^{-1})}{n}+\sum_{N=-n+1}^{-1}\frac{1}{|N|}\right]\to\infty
\quad \mbox{as} \quad n\to\infty.
$$
Thus, $f(x)$ cannot be continuous at $x=0$. Theorem \ref{l1} is
proved. \rule{2mm}{2mm}

\subsection{Properties of solutions of $D^{\alpha}+I=\delta$.}

Let us consider an equation
\begin{equation}\label{e12}
D^{\alpha}h+h=\delta_{x_k}, \qquad h\in{L_2(\mathbb{Q}_p)}, \quad
x_k\in\mathbb{Q}_p, \quad \alpha>0,
\end{equation}
where $D^{\alpha}: {L_2(\mathbb{Q}_p)} \to
\mathcal{D}'(\mathbb{Q}_p)$ is understood in the distribution sense.

It is known \cite{KO} that  Eq. (\ref{e12}) has a unique solution
$h=h_k\in{L_2(\mathbb{Q}_p)}$  for $\alpha>1/2$ and has no solutions
belonging to $L_2(\mathbb{Q}_p)$ for $\alpha\leq{1/2}$. The next
statement continues the investigation of $h_k$.

\begin{lelele}\label{pifpaf}
The solution $h_k$ of (\ref{e12}) is a function continuous on
$\mathbb{Q}_p$ when $\alpha>1$ and continuous on
$\mathbb{Q}_p\backslash\{x_k\}$ when $1/2<\alpha\le 1$.
\end{lelele}

{\it Proof.}
 Reasoning as in the proof of (\cite[Lemma 3.7]{KO}), where the
basis of Vladimirov eigenfunctions was used, we establish the
expansion of $\delta_{x_k}$ in terms of the $p$-adic wavelet basis.

Let $u\in\mathcal{D}(D^{\alpha})$. By analogy with the proof of
Theorem \ref{l1} we expand $u$ in an uniformly convergent series
with respect to the complex-conjugated $p$-adic wavelet basis
$\{\overline{\psi_{Nj\epsilon}}\}$. Since
$\{\overline{\psi_{Nj\epsilon}}\}$ are continuous functions on
$\mathbb{Q}_p$  we can write:
$u(x_k)=\sum_{N=-\infty}^{\infty}\sum_{j=1}^{p-1}\sum_{\epsilon}(u,\overline{\psi_{Nj\epsilon}})\overline{\psi_{Nj\epsilon}}(x_k)
$ for $x=x_k$.

Consider
$$
\overline{\psi_{Nj\epsilon}}(x_k)=p^{-N/2}\overline{\chi(p^{N-1}jx_k)}\Omega(|p^Nx_k-\epsilon|_p)=p^{-N/2}\chi(-p^{N-1}jx_k)\Omega(|p^Nx_k-\epsilon|_p).
$$

Obviously, $\overline{\psi_{Nj\epsilon}}(x_k)\not=0 \iff
|p^Nx_k-\epsilon|_p\leq{1}$. Here
$\epsilon\in\mathbb{Q}_p/\mathbb{Z}_p$ and  hence, $|\epsilon|_p>1$
for $\epsilon\ne 0$. It follows from the strong triangle inequality
and the condition $\epsilon\in\mathbb{Q}_p/\mathbb{Z}_p$ that
$|p^Nx_k-\epsilon|_p\leq{1} \iff \epsilon=\{p^Nx_k\}_p$ (if
$\epsilon\ne 0$). Moreover, if $\epsilon=0$, then condition
$|p^{N}x_k|_p\leq{1}$ implies $\{p^Nx_k\}_p=0$. Combining these two
cases we arrive at the conclusion that
$$
\overline{\psi_{Nj\epsilon}}(x_k)= \left\{ \begin{array}{ll}
0, &  \epsilon\ne\{p^Nx_k\}_p  \\
p^{-N/2}\chi(-p^{N-1}jx_k), &  \epsilon=\{p^Nx_k\}_p
\end{array}
\right.
$$
But then
\begin{eqnarray}\label{oh1}
<\delta_{x_k},
u>=u(x_k) &=& \sum_{N=-\infty}^{\infty}\sum_{j=1}^{p-1}p^{-N/2}\chi(-p^{N-1}jx_k)\big(u,\overline{\psi_{Nj\{p^Nx_k\}_p}}\big) \\
&=&
\sum_{N=-\infty}^{\infty}\sum_{j=1}^{p-1}p^{-N/2}\chi(-p^{N-1}jx_k)<\psi_{Nj\{p^Nx_k\}_p},u>
\nonumber
\end{eqnarray}

Since $\mathcal{D}(\mathbb{Q}_p)\subset\mathcal{D}(D^{\alpha})$, the
equality (\ref{oh1}) means that
\begin{equation}\label{ee13}
\delta_{x_k}=\sum_{N=-\infty}^{\infty}\sum_{j=1}^{p-1}p^{-N/2}\chi(-p^{N-1}jx_k)\psi_{Nj\{p^Nx_k\}_p},
\end{equation}
where the series converges  in  $\mathcal{D}'(\mathbb{Q}_p)$.

Suppose that a function $h_k\in{L_2(\mathbb{Q}_p)}$ is represented
as a convergent series in $L_2(\mathbb{Q}_p)$:
$$
h_k(x)=\sum_{N=-\infty}^{\infty}\sum_{j=1}^{p-1}\sum_{\epsilon}c_{Nj\epsilon}\psi_{Nj\epsilon}(x).
$$
Applying the operator $D^{\alpha}+I$ termwise, we get a series
\begin{equation}\label{ee14}
D^{\alpha}h_k+h_k=\sum_{N=-\infty}^{\infty}\sum_{j=1}^{p-1}\sum_{\epsilon}c_{Nj\epsilon}\big(1+p^{\alpha(1-N)}\big)\psi_{Nj\epsilon},
\end{equation}
converging in $\mathcal{D}'$ (since
$D^{\alpha}\mathcal{D}(\mathbb{Q}_p)\subset{L_2}(\mathbb{Q}_p)$).
Comparing the terms of (\ref{ee13}) and (\ref{ee14}) gives
$$
c_{Nj\epsilon}=\left\{
\begin{array}{ll}
  0, & \epsilon\ne\{p^Nx_k\}_p \\
  p^{-N/2}\chi(-p^{N-1}jx_k)\big[p^{\alpha(1-N)}+1\big]^{-1}, &
  \epsilon=\{p^Nx_k\}_p
\end{array}
\right.
$$
Thus,
\begin{equation}\label{ee15}
h_k(x)=\sum_{N=-\infty}^{\infty}\sum_{j=1}^{p-1}p^{-N/2}\chi(-p^{N-1}jx_k)\big[p^{\alpha(1-N)}+1\big]^{-1}\psi_{Nj\{p^Nx_k\}_p}(x).
\end{equation}

Let us show that the series (\ref{ee15}) is uniformly convergent on
$\mathbb{Q}_p$ for $\alpha>1$ and is uniformly convergent on any
ball not containing $x_k$ for $1/2<\alpha\le 1$.

Indeed, by virtue of (\ref{a8}) the general term of (\ref{ee15})
does not exceed
\begin{equation}\label{fa}
p^{-N}\big[p^{\alpha(1-N)}+1\big]^{-1}\le p^{-N}.
\end{equation}
Hence, the subseries of (\ref{ee15}) formed by terms with $N\geq{0}$
converges uniformly.

For $N<0$ the general term of (\ref{ee15}) does not exceed
$$
p^{-N}\big[p^{\alpha(1-N)}+1\big]^{-1}\le
\frac{1}{p^{\alpha}}p^{-N(1-\alpha)}.
$$
The obtained estimate implies that for $\alpha>1$ the subseries of
(\ref{ee15}) formed by terms with $N<{0}$ also converges uniformly.
So, the series (\ref{ee15}) converges uniformly for $\alpha>1$. This
proves the assertion of Lemma \ref{pifpaf} for $\alpha>1$ (since
$\psi_{Nj\epsilon}$ are continuous on $\mathbb{Q}_p$).

Let $B_\gamma(a)$ be a ball such that $x_k\not\in{B_\gamma(a)}$. To
prove Lemma \ref{pifpaf} for $1/2<\alpha\le 1$ it suffices to verify
that the restriction of (\ref{ee15}) onto $B_\gamma(a)$ contains
finite number of terms with negative parameter $N<0$.

Indeed, it follows from the strong triangle inequality and the
definitions of $\{\cdot\}_p$ and $\Omega(\cdot)$ (see (\ref{at1}))
that
\begin{equation}\label{as6}
\Omega(|p^Nx-\{p^Nx_k\}_p|_p)=\Omega(|p^Nx-p^Nx_k|_p).
\end{equation}
Hence, the restriction of $\psi_{Nj\{p^Nx_k\}_p}(x)$ onto
$B_\gamma(a)$
 is equal to $0$ if $|x-x_k|_p>p^N$ for all $x\in{B_\gamma(a)}$. Since
$|x-x_k|_p>p^\gamma$, the relation
$\psi_{Nj\{p^Nx_k\}_p}(x)\equiv{0}$ ($\forall{x}\in{B}_\gamma(a)$)
holds for all $N\leq{\gamma}$. Using the estimation (\ref{fa}) we
arrive at the conclusion that the series (\ref{ee15}) converges
uniformly for any ball
$B_\gamma(a)\subset\mathbb{Q}_p\setminus\{x_k\}$. Lemma \ref{pifpaf}
is proved. \rule{2mm}{2mm}

{\bf Remark.} The solution $h_k(x)$ of (\ref{e12}) constructed in
Lemma \ref{pifpaf} is a real-valued function. This fact can be
obtained directly from the expansion (\ref{ee15}). Another way to
establish it is based on the invariance of the space
$\mathcal{D}(\mathbb{Q}_p)$ and the operator $D^{\alpha}$ with
respect to the complex conjugation. Combining these properties with
the uniqueness of the solution of $D^{\alpha}+I=\delta_{x_k}$ in
$L_2(\mathbb{Q}_p)$, we get $\overline{h}_k(x)=h_k(x)$.

\begin{ccc}\label{c1}
Let the index $\alpha>1/2$ and points $x_1, \ldots,
x_n\in\mathbb{Q}_p$ be fixed and let $\mathrm{Sp}\{h_k\}_{1}^n$ be
the linear span of solutions $h_k$ ($1\leq{k}\leq{n}$) of
(\ref{e12}). Then
$\mathrm{Sp}\{h_k\}_{1}^n\cap\mathcal{D}(D^{\alpha/2})=\{0\}$ for
$1/2<\alpha\leq{1}$ and
$\mathrm{Sp}\{h_k\}_{1}^n\subset\mathcal{D}({D^{\alpha/2}})$ for
$\alpha>1$.
\end{ccc}

{\it Proof.} The solution $h_k$ of (\ref{e12}) is determined by
(\ref{ee15}). Taking the expansion (\ref{ee15}) and the ``semigroup
property''
\begin{equation}\label{asa1}
D^{\alpha_1}D^{\alpha_2}=D^{\alpha_1+\alpha_2}, \qquad \alpha_1,
\alpha_2>0
\end{equation}
of $D^{\alpha}$ into account, it is easy to see that
$h_k\in\mathcal{D}(D^{\alpha/2})$ if and only if the following
series converge in $L_2(\mathbb{Q}_p)$:
\begin{eqnarray*}
{}& \sum_{N=1}^{\infty}\sum_{j=1}^{p-1}p^{-N/2}\chi(-p^{N-1}jx_k)\big[p^{\alpha(1-N)}+1\big]^{-1}p^{\frac{\alpha}{2}(1-N)}\psi_{Nj\{p^Nx_k\}_p}+\\
&\sum_{N=-\infty}^{0}\sum_{j=1}^{p-1}p^{-N/2}\chi(-p^{N-1}jx_k)\big[p^{\alpha(1-N)}+1\big]^{-1}p^{\frac{\alpha}{2}(1-N)}\psi_{Nj\{p^Nx_k\}_p}
\end{eqnarray*}
(if the limit exists then it coincides with $D^{\alpha/2}h_k$). For
the general term of the first series we have
$$
\big|p^{-N/2}p^{\frac{\alpha}{2}(1-N)}\chi(-p^{N-1}jx_k)\big[p^{\alpha(1-N)}+1\big]^{-1}\big|^2\le
p^{-N(\alpha+1)+\alpha}, \quad N\geq{1}
$$
that implies its convergence in $L_2(\mathbb{Q}_p)$ for any
$\alpha>1/2$.

Similarly, the general term of the second series can be estimated as
follows:
$$
\big|p^{-N/2}p^{\frac{\alpha}{2}(1-N)}\chi(-p^{N-1}jx_k)\big[p^{\alpha(1-N)}+1\big]^{-1}\big|^2\le
Cp^{(\alpha-1)N}, \quad N\leq{0}.
$$
Obviously this series converges for $\alpha>1$. Thus
$\mathrm{Sp}\{h_k\}_{1}^n\subset\mathcal{D}(D^{\alpha/2})$ for
$\alpha>1$.

Since $p^{\alpha(1-N)}+1\le 2p^{\alpha(1-N)}$ for $N\le 0$, we can
estimate from below the general term of the second series
\begin{equation}\label{as2}
\frac{1}{4p^{\alpha}}p^{(\alpha-1)N}\leq\big|p^{-N/2}p^{\frac{\alpha}{2}(1-N)}\chi(-p^{N-1}jx_k)\big[p^{\alpha(1-N)}+1\big]^{-1}\big|^2
 \quad (N\leq{0})
\end{equation}
that implies its divergence in $L_2(\mathbb{Q}_p)$ for $\alpha\le
1$.

Thus $h_k\not\in\mathcal{D}({D^{\alpha/2}})$. From this, taking into
account that the estimate (\ref{as2}) does not depend on the choice
of $h_k$ and the functions $\{\psi_{Nj\{p^Nx_k\}_p}(x)\}$ ($N<0$) of
the basis $\{\psi_{Nj\epsilon}(x)\}$ corresponding to $h_k$
($1\leq{k}\leq{n}$) in (\ref{ee15}) are different for sufficiently
small negative indexes $N$, we conclude that
$\mathrm{Sp}\{h_k\}_{1}^n\cap\mathcal{D}(D^{\alpha/2})=\{0\}$ for
$1/2<\alpha\leq{1}$. Corollary \ref{c1} is proved. \rule{2mm}{2mm}

\setcounter{equation}{0}
\section{Operator $D^{\alpha}$ with Point Interactions}
\subsection{The Friedrichs extension.}
Let $\mathfrak{H}_2\subset\mathfrak{H}_1\subset{L_2}(\mathbb{Q}_p)
 \subset\mathfrak{H}_{-1}\subset\mathfrak{H}_{-2}$
 be the standard scale of Hilbert spaces ($A$-scale) associated
with the positive self-adjoint operator $A=D^{\alpha}$. Here
${\mathfrak H}_s=\mathcal{D}(A^{s/2})$, $s=1,2$, with the norm
$\|u\|_s=\|(D^{\alpha}+I)^{s/2}u\|$ and ${\mathfrak H}_{-s}$ are the
completion of ${L_2}(\mathbb{Q}_p)$ with respect to the norm
$\|u\|_{-s}$ (see \cite{AL1}, \cite{Ber2} for details).

Recalling that $h_k(x)$ is a real-valued function and employing
(\ref{oh1}), (\ref{ee14}) (with $u$ and
$\overline{\psi_{Nj\epsilon}}$ instead of $h_k$ and
$\psi_{Nj\epsilon}$, respectively),
 and (\ref{ee15}),
we get
\begin{equation}\label{ada5}
<\delta_{x_k},u>=u(x_k)=((D^{\alpha}+I)u, h_k)_{L_2(\mathbb{Q}_p)},
\quad  \forall{u}\in{\mathcal{D}}(D^{\alpha}), \
 x_k\in\mathbb{Q}_p.
\end{equation}

Thus, the Dirac delta function $\delta_{x_k}$ is well posed on
${\mathfrak H}_2=\mathcal{D}(D^{\alpha})$ and
$\delta_{x_k}\in{\mathfrak H}_{-2}$ for $\alpha>1/2$.

Let us fix points $x_1, \ldots, x_n$ ($n<\infty$) from
$\mathbb{Q}_p$ and consider a positive symmetric operator
\begin{equation}\label{ee4}
A_{\mathrm{sym}}={D^{\alpha}}\upharpoonright_{\mathcal{D}}, \quad
\mathcal{D}=\{u\in\mathcal{D}(D^{\alpha}) \ | \
u(x_1)=\ldots=u(x_n)=0\}.
\end{equation}

By Theorem \ref{l1} the formula (\ref{ee4}) is well-posed for
$\alpha>1/2$. In this case, (\ref{ada5}) implies that
$A_{\mathrm{sym}}$ is a closed densely defined operator in
$L_2(\mathbb{Q}_p)$ and its defect subspace
$\mathcal{H}=\ker(A_{\mathrm{sym}}^*+I)$ coincides with the linear
span of $\{h_k\}_{k=1}^{n}$. Hence, the deficiency index of
$A_{\mathrm{sym}}$ is equal to $(n,n)$.

It is clear that the domain of the adjoint $A_{\mathrm{sym}}^*$ has
the form ${\mathcal
D}(A_{\mathrm{sym}}^*)={\mathcal{D}}(D^{\alpha})\dot{+}\mathcal{H}$
and
\begin{equation}\label{ee5}
A_{\mathrm{sym}}^*f=A_{\mathrm{sym}}^*(u+h)=D^{\alpha}u-h, \qquad
\forall{f}=u+h\in{\mathcal D}(A_{\mathrm{sym}}^*)
\end{equation}
($u\in\mathcal{D}(D^{\alpha}), \ h\in\mathcal{H}$).

\begin{ppp}\label{p22}
Let $A_F$ be the Friedrichs extension of $A_{\mathrm{sym}}$. Then
$A_F=D^{\alpha}$ when $1/2<\alpha\leq{1}$ and
$$
A_F=A_{\mathrm{sym}}^*\upharpoonright_{\mathcal{D}(A_F)}, \quad
\mathcal{D}(A_F)=\{f(x)\in\mathcal{D}(A_{\mathrm{sym}}^*) \ | \
f(x_1)=\ldots={f(x_n)=0}\}
$$
when $\alpha>1$.
\end{ppp}

{\it Proof.}  It follows from (\ref{asa1}) that
$\mathfrak{H}_1=\mathcal{D}(D^{\alpha/2})$. This relation and
Corollary \ref{c1} mean that $\mathcal{H}\subset\mathfrak{H}_{1}$ \
($\alpha>1$) and $\mathcal{H}\cap{\mathfrak{H}_{1}}=\{0\}$ \
($1/2<\alpha\leq{1}$).

After such a preparation work, the proof is a direct consequence of
some `folk-lore' results of the extension theory. For the
convenience of the reader some principal stages are repeated below.

First of all, we recall that the Friedrichs extension $A_F$ of
$A_{\mathrm{sym}}$ is defined as the restriction of the adjoint
$A_{\mathrm{sym}}^*$ onto
$\mathcal{D}(A_F)=\mathcal{D}\cap\mathcal{D}(A_{\mathrm{sym}}^*)$,
where $\mathcal{D}$ is the completion of
$\mathcal{D}(A_{\mathrm{sym}})$ in the Hilbert space
$\mathfrak{H}_{1}$.

Using the obvious equality
$\mathfrak{H}_{1}=\mathcal{D}\oplus_1{\mathcal{H}}'$ (here
${\mathcal{H}}'={\mathcal{H}}\cap\mathfrak{H}_{1}$ and $\oplus_1$
denotes the orthogonal sum in $\mathfrak{H}_{1}$), we describe
$\mathcal{D}(A_F)$ as follows:
$$
\mathcal{D}(A_F)=\{f\in\mathfrak{H}_1\cap\mathcal{D}(A_{\mathrm{sym}}^*)
\ | \ ((D^{\alpha}+I)^{1/2}f,
(D^{\alpha}+I)^{1/2}h')_{L_2(\mathbb{Q}_p)}=0,
\forall{h'}\in\mathcal{H}'\}.
$$

If $\mathcal{H}\cap{\mathfrak{H}_{1}}=\{0\}$ (the case
$1/2<\alpha\leq{1}$), then $\mathcal{H}'=\{0\}$ and
$\mathcal{D}(A_F)=\mathfrak{H}_{1}\cap\mathcal{D}(A_{\mathrm{sym}}^*)=\mathcal{D}(D^{\alpha})$.
Thus $A_F=D^{\alpha}$.

If $\mathcal{H}\subset\mathfrak{H}_{1}$ (the case $\alpha>1$), then
$\mathcal{H}'=\mathcal{H}$,
$\mathcal{D}(A_{\mathrm{sym}}^*)\subset\mathfrak{H}_1$ and
$$
\mathcal{D}(A_F)=\{f\in\mathcal{D}(A_{\mathrm{sym}}^*) \ | \
((D^{\alpha}+I)^{1/2}f,
(D^{\alpha}+I)^{1/2}{h_k})_{L_2(\mathbb{Q}_p)}=0, \
1\leq{k}\leq{n}\}.
$$

Repeating the same arguments as in the proof of (\ref{ada5}) it is
easy to see that $((D^{\alpha}+I)^{1/2}f,
(D^{\alpha}+I)^{1/2}{h_k})_{L_2(\mathbb{Q}_p)}=f(x_k)$. Proposition
\ref{p22} is proved. \rule{2mm}{2mm}

\subsection{Operator realizations of $D^{\alpha}+V_Y$ in $L_2(\mathbb{Q}_p)$.}
In the additive singular perturbations theory, the algorithm of the
determination of operator realizations of finite rank point
perturbations of $D^{\alpha}$ is determined by the general
expression
\begin{equation}\label{eee1}
A_Y=D^{\alpha}+V_Y, \qquad V_Y=\sum_{i,j=1}^nb_{ij}<\delta_{x_j},
\cdot>\delta_{x_i}, \quad   b_{ij}\in\mathbb{C},
\end{equation}
$Y=\{x_1,\ldots,x_n\}$ is well known \cite{AL1} and it is based on
the construction of some extension (regularization)
${A}_{Y\mathrm{reg}}:=D^\alpha+V_{Y\mathrm{reg}}$ of (\ref{eee1})
onto the domain ${\mathcal
D}(A_{\mathrm{sym}}^*)={\mathcal{D}}(D^{\alpha})\dot{+}\mathcal{H}$.

The  $L_2(\mathbb{Q}_p)$-part
\begin{equation}\label{les40}
\widetilde{A}={A}_{Y\mathrm{reg}}\upharpoonright_{\mathcal{D}(\widetilde{A})},
\ \ \ \
\mathcal{D}(\widetilde{A})=\{f\in\mathcal{D}(A_{\mathrm{sym}}^*) \ |
\ {A}_{Y\mathrm{reg}}f\in{L_2(\mathbb{Q}_p)}\}
\end{equation}
of the regularization ${A}_{Y\mathrm{reg}}$ is called the {\it
operator realization} of $D^\alpha+V_Y$ in $L_2(\mathbb{Q}_p)$.

Since the action of $D^{\alpha}$ on elements of $\mathcal{H}$ is
defined by (\ref{e12}), the regularization ${A}_{Y\mathrm{reg}}$
depends on the determination of $V_{Y\mathrm{reg}}$.

If $\alpha>1$, the singular potential
$V_Y=\sum_{i,j=1}^nb_{ij}<\delta_{x_j}, \cdot>\delta_{x_i}$ is form
bounded (since all ${h_k}\in\mathfrak{H}_{1}$ and hence, all
$\delta_{x_k}\in\mathfrak{H}_{-1}$). In this case,
$\mathcal{D}(A_{\mathrm{sym}}^*)\subset\mathfrak{H}_{1}$ consists of
continuous functions on $\mathbb{Q}_p$ (Lemma \ref{pifpaf}) and the
delta functions $\delta_{x_k}$ are uniquely determined on elements
$f\in\mathcal{D}(A_{\mathrm{sym}}^*)$ by continuity (cf.
(\ref{ada5}))
\begin{equation}\label{as4}
<\delta_{x_k},f>=((D^{\alpha}+I)^{1/2}f,
(D^{\alpha}+I)^{1/2}{h_k})_{L_2(\mathbb{Q}_p)}=f(x_k).
\end{equation}

Thus, for $\alpha>1$, the regularization ${A}_{Y\mathrm{reg}}$ is
uniquely defined and formula (\ref{les40}) provides a unique
operator realization of (\ref{eee1}) in $L_2(\mathbb{Q}_p)$
corresponding to a fixed singular potential $V_Y$.

The case $1/2<\alpha\leq{1}$ is more complicated, because
$\delta_{x_k}$ cannot be extended onto
$\mathcal{D}(A_{\mathrm{sym}}^*)$ by continuity.

Since any function $f\in{\mathcal
D}(A_{\mathrm{sym}}^*)={\mathcal{D}}(D^{\alpha})\dot{+}\mathcal{H}$
admits a decomposition $f=u+\sum_{j=1}^n{c}_j{h}_j$ \ \
($u\in{\mathcal{D}}(D^{\alpha}), \  c_i\in\mathbb{C})$, the
extension of $\delta_{x_k}$ on ${\mathcal D}(A_{\mathrm{sym}}^*)$ is
well determined if the entries $r_{kj}=<\delta_{x_k}, h_j>$ of the
matrix $\mathcal{R}=(r_{kj})_{k,j=1}^n$ are known. In this case, the
extended delta-function $\delta_{x_k}$ acts on functions
$f\in\mathcal{D}(A_{\mathrm{sym}}^*)$ by the rule
 \begin{equation}\label{k23}
<\delta_{x_k}, f>=u(x_k)+c_1r_{k1}+\ldots{+}c_nr_{kn}, \qquad
1\leq{k}\leq{n}.
 \end{equation}
(We preserve the same notation $\delta_{x_k}$ for the extension.)

Since $\mathcal{H}\cap{\mathfrak{H}_{1}}=\{0\}$, the system
$\{\delta_{x_k}\}_{k=1}^n$ is ${\mathfrak{H}_{-1}}$-independent
(i.e., its linear span
$\mathrm{Sp}\{\delta_{x_k}\}_{1}^n\cap\mathfrak{H}_{-1}=\{0\}$).
Therefore, the natural restrictions on the choice of $r_{kj}$ in
(\ref{k23}) induced by the fact that a functional $<\phi, \cdot>$
where
$\phi\in\mathrm{Sp}\{\delta_{x_k}\}_{1}^n\cap\mathfrak{H}_{-1}$
admits a natural extension by continuity onto
$\mathfrak{H}_{1}\cap\mathcal{D}(A_{\mathrm{sym}}^*)$  do not appear
in our case (see \cite{AL1} for details). This means that, in
general, any Hermitian matrix $\mathcal{R}=(r_{kj})_{k,j=1}^n$ can
be used for the determination of the extended functionals
$<\delta_{x_k}, \cdot>$ in (\ref{k23}).

One of the possible approaches to the definition of $r_{kj}$ deals
with the fact that the functions $h_j(x)$ turn out to be continuous
at the point $x=x_k$ if $j\not=k$ (see Lemma \ref{pifpaf}). In view
of this, it is natural to assume that
\begin{equation}\label{as12}
r_{kj}=<\delta_{x_k}, h_j>=h_j(x_k), \qquad j\not=k.
\end{equation}

However this formula cannot be used for the definition of $r_{kk}$
because the substitution of $x_k$ for $x$ in (\ref{ee15}) leads to
the formal equality
\begin{equation}\label{as13}
h_k(x_k)=(p-1)\sum_{N=-\infty}^{\infty}\frac{p^{-N}}{p^{\alpha(1-N)}+1}
\end{equation}
with a divergent series at the right-hand side. Note that this
series do not depend on $k$. For this reason, some choice of a real
number $r=r_{kk}$ ($1\leq{k}\leq{n}$) can be interpreted as a
certain regularization of
$\sum_{-\infty}^{\infty}\frac{p^{-N}}{p^{\alpha(1-N)}+1}$.

It follows from (\ref{ee15}), (\ref{as6}), and (\ref{as12}) that
$r_{kj}=\overline{r}_{jk}$. Hence, the matrix
$\mathcal{R}=(r_{kj})_{k,j=1}^n$ constructed in such a way is
Hermitian.

It should be noted that if we will use (\ref{k23}) instead of the
direct formula (\ref{as4}) for the definition of extensions
$<\delta_{x_k}, \cdot>$ in the case $\alpha>1$, then we arrive at
just the same form of the matrix $\mathcal{R}$. The difference only
is in the convergency of the series in (\ref{as13}) for $\alpha>1$
and hence, $r_{kk}=h_k(x_k)$.

\subsection{Description of operator realizations.}

Let $\eta$ be an invertible bounded self-adjoint operator in
$L_2(\mathbb{Q}_p)$.

An operator $A$ is called $\eta$-{\it self-adjoint} in
$L_2(\mathbb{Q}_p)$ if $A^{*}={\eta}A{\eta^{-1}}$, where $A^{*}$
denotes the adjoint of $A$ \cite{AZ}. Obviously, self-adjoint
operators are a particular case of $\eta$-self-adjoint ones for
$\eta=I$. In this case, we will use notation `self-adjoint' instead
of `$I$-self-adjoint'.

We are going to describe  $\eta$-self-adjoint operator realizations
$\widetilde{A}$ (see (\ref{les40})) of $D^{\alpha}+V_Y$ in
$L_2(\mathbb{Q}_p)$.

To do this, we determine linear mappings $\Gamma_i :{\mathcal
D}(A_{\mathrm{sym}}^*)\to{\mathbb C}^n$ ($i=0,1$):
\begin{equation}\label{k9}
\Gamma_0f=\left(\begin{array}{c}
 <\delta_{x_1}, f> \\
 \vdots \\
 <\delta_{x_n}, f>
\end{array}\right), \quad \Gamma_1f=-\left(\begin{array}{c}
 c_1 \\
 \vdots \\
c_n
\end{array}\right), \quad
\forall{f}=u+\sum_{i=1}^n{c}_i{h}_i\in{\mathcal
D}(A_{\mathrm{sym}}^*).
\end{equation}

In what follows we will assume that
\begin{equation}\label{ne2}
D^{\alpha}\eta={\eta}D^{\alpha} \qquad  \mbox{and} \qquad
 \eta:\mathcal{H}\to{\mathcal H}.
 \end{equation}

By the second relation in (\ref{ne2}), the action of $\eta$ on
elements of ${\mathcal H}$ can be described with the help of a
matrix $\mathcal{Y}=({y}_{ij})^n_{i,j=1}$, i.e.,
\begin{equation}\label{as17}
\eta\sum_{i=1}^n{c}_i{h}_i=({h}_1,\ldots,
{h}_n)\mathcal{Y}(c_1,\ldots,c_n)^{\mathrm{t}} \qquad
(c_i\in\mathbb{C}),
\end{equation}
where the upper index ${}^{\mathrm{t}}$ denotes the operation of
transposition. Since, in general, the basis $\{{h}_i\}_{i=1}^{n}$ of
$\mathcal{H}$ is not orthogonal, the matrix $\mathcal{Y}$ is not
Hermitian ($\mathcal{Y}\not=\overline{\mathcal{Y}}^{\mathrm{t}}$).

\begin{lelele}\label{as1}
If $\alpha>1$, then
$$
\Gamma_0{{\eta}f}=\overline{\mathcal{Y}}^{\mathrm{t}}\Gamma_0f
\qquad \mbox{and} \qquad \Gamma_1{{\eta}f}=\mathcal{Y}\Gamma_1f
\quad (\forall{f}\in{\mathcal D}(A_{\mathrm{sym}}^*)).
$$

These equalities also hold for $1/2<\alpha\leq{1}$ if
$\mathcal{RY}=\overline{\mathcal{Y}}^{\mathrm{t}}\mathcal{R}$, where
the matrix $\mathcal{R}$ determines the extended functionals
$<\delta_{x_k}, \cdot>$ in (\ref{k23}).
\end{lelele}

{\it Proof.} Let $f=u+\sum_{i=1}c_i{h}_i\in{\mathcal
D}(A_{\mathrm{sym}}^*)$. By (\ref{as17})
\begin{equation}\label{as16}
\eta{f}=\eta{u}+(h_1,\ldots,
h_n)\mathcal{Y}(c_1,\ldots,c_n)^{\mathrm{t}},
\end{equation}
where $\eta{u}\in{\mathcal{D}}(D^{\alpha})$ (see the first relation
in (\ref{ne2})). In view of (\ref{k9}),
$$
{\Gamma}_1\eta{f}=-\mathcal{Y}(c_1,\ldots,c_n)^{\mathrm{t}}=
\mathcal{Y}{\Gamma}_1{f}.
$$

It follows from the first relation in (\ref{ne2}) that
$\eta(D^\alpha+I)^{1/2}=(D^\alpha+I)^{1/2}\eta$. Taking this
equality into account, we deduce from (\ref{as4})
$$
\begin{array}{l}
 <\delta_{x_k}, \eta{f}>=((D^\alpha+I)^{1/2}f, (D^\alpha+I)^{1/2}{\eta}h_k)_{L_2(\mathbb{Q}_p)}=
\\
=(\overline{y}_{1k},\ldots, \overline{y}_{nk})(<\delta_{x_1}, f>,
\ldots, <\delta_{x_n},f>)^{\mathrm{t}}
\end{array}
$$
that implies
${\Gamma}_0\eta{f}=\overline{\mathcal{Y}}^{\mathrm{t}}{\Gamma}_0{f}$
for $\alpha>1$.

Similar arguments with the employing (\ref{k23}), (\ref{as16}), and
$\mathcal{RY}=\overline{\mathcal{Y}}^{\mathrm{t}}\mathcal{R}$ give
$$
\begin{array}{l}
\Gamma_0\eta{f}=\Gamma_0\eta{u}+\mathcal{RY}(c_1,\ldots,c_n)^{\mathrm{t}}=
\overline{\mathcal{Y}}^{\mathrm{t}}{\Gamma}_0{u}+\overline{\mathcal{Y}}^{\mathrm{t}}\mathcal{R}(c_1,\ldots,c_n)^{\mathrm{t}}=
\\
\overline{\mathcal{Y}}^{\mathrm{t}}{\Gamma}_0[u+({h}_1,\ldots,
{h}_n)(c_1,\ldots,c_n)^{\mathrm{t}}]=\overline{\mathcal{Y}}^{\mathrm{t}}{\Gamma}_0f
\end{array}
$$
for $1/2<\alpha\leq{1}$. Lemma \ref{as1} is proved. \rule{2mm}{2mm}

\begin{ttt}\label{ss1}
Let  $\widetilde{A}$ be the operator realization of $D^{\alpha}+V_Y$
defined by (\ref{les40}). Then $\widetilde{A}$ coincides with the
operator
\begin{equation}\label{k4141}
{A}_{\mathcal{B}}=A_{\mathrm{sym}}^*\upharpoonright{\mathcal{D}(A_{\mathcal{B}})},
\quad
\mathcal{D}(A_{\mathcal{B}})=\{f\in\mathcal{D}(A_{\mathrm{sym}}^*) \
| \ {\mathcal B}\Gamma_0f=\Gamma_1f\},
\end{equation}
where ${\mathcal{B}}=(b_{ij})_{i,j=1}^n$ is the coefficient matrix
of the singular potential $V_Y$.

The operator ${A}_{\mathcal{B}}$ is self-adjoint if and only if the
matrix $\mathcal{B}$ is Hermitian.

If $\eta$ satisfy (\ref{ne2}) and $\alpha>1$, then
${A}_{\mathcal{B}}$ is $\eta$-self-adjoint if and only if the matrix
$\mathcal{YB}$ is Hermitian. This statement is also true for the
case $1/2<\alpha\leq{1}$ under the additional condition that the
matrix $\mathcal{RY}$ is Hermitian, where $\mathcal{R}$ determines
the extended functionals $<\delta_{x_k}, \cdot>$ in (\ref{k23}).
\end{ttt}

{\it Proof.} It follows from (\ref{e12}), (\ref{ee5}), and
(\ref{k9}) that
$$
{A}_{V\mathrm{reg}}f=A_{\mathrm{sym}}^*f+(\delta_{x_1},\ldots,\delta_{x_n})(
\mathcal{B}\Gamma_0f-\Gamma_1f), \quad f\in{\mathcal
D}(A_{\mathrm{sym}}^*)
$$
This equality and (\ref{les40}) mean that the operator realization
$\widetilde{A}$ of $D^{\alpha}+V_Y$ coincides with the operator
$A_{\mathcal{B}}$ determined by (\ref{k4141}).

It is known (see, e.g., \cite{AKN}, \cite{DHS}) that the triple
$(\mathbb{C}^n, \Gamma_0, \Gamma_1)$, where $\Gamma_i$ are defined
by (\ref{k9}), is a boundary value space (BVS) of
$A_{\mathrm{sym}}$. This means that the abstract Green identity
\begin{equation}\label{tat15}
({A_{\mathrm{sym}}^*}f, g)-(f, {A_{\mathrm{sym}}^*}g)=(\Gamma_1f,
\Gamma_0g)_{\mathbb{C}^n}-(\Gamma_0f, \Gamma_1g)_{\mathbb{C}^n}, \ \
\ \ f, g\in\mathcal{D}({A_{\mathrm{sym}}^*})
\end{equation}
is satisfied and the map $(\Gamma_0,
\Gamma_1):\mathcal{D}({A_{\mathrm{sym}}^*})\to\mathbb{C}^n\oplus\mathbb{C}^n$
is surjective.

It follows from the general results of the BVS-theory \cite{Gor},
\cite{GGK}, \cite{KK} that the operator ${A}_{\mathcal{B}}$
determined by (\ref{k4141}) is self-adjoint $\iff$ the matrix
${\mathcal B}$ is Hermitian.

Conditions (\ref{ne2}) imposed on $\eta$ ensure the commutativity of
$\eta$ with $A_{\mathrm{sym}}$ and $A_{\mathrm{sym}}^*$, i.e.,
\begin{equation}\label{a10}
\eta{A_{\mathrm{sym}}}=A_{\mathrm{sym}}\eta, \qquad
\eta{A_{\mathrm{sym}}^*}=A_{\mathrm{sym}}^*\eta.
\end{equation}

Relations (\ref{a10}) and the definition of $\eta$-self-adjoint
operators imply that ${A}_{\mathcal{B}}$ is $\eta$-self-adjoint
$\iff$ $\eta{A}_{\mathcal{B}}$ is a self-adjoint extension of the
symmetric operator $F_{\mathrm{sym}}=\eta{A_{\mathrm{sym}}}$.

Thus, the description of $\eta$-self-adjoint operators is reduced to
the similar problem for self-adjoint ones.

It immediately follows from Lemma \ref{as1} and relations
(\ref{tat15}), (\ref{a10}) that the triple $(\mathbb{C}^n, \Gamma_0,
\mathcal{Y}\Gamma_1)$ is a BVS for the symmetric operator
$F_{\mathrm{sym}}$. In this BVS, the operator
$\eta{A}_{\mathcal{B}}$ is described by the formula (cf.
(\ref{k4141})):
$$
\eta{A}_{\mathcal{B}}={\eta}A_{\mathrm{sym}}^*\upharpoonright_{\mathcal{D}({\eta}A_{\mathcal{B}})},
\quad
\mathcal{D}({\eta}A_{\mathcal{B}})=\{f\in\mathcal{D}(A_{\mathrm{sym}}^*)
\ | \ {\mathcal YB}\Gamma_0f={\mathcal Y}\Gamma_1f\}
$$
that completes the proof of Theorem \ref{ss1}. \rule{2mm}{2mm}


 \end{document}